\documentclass[a4paper,12pt]{article}

\usepackage{amsmath}
\usepackage[psamsfonts]{amssymb}
\usepackage{rsfs}
\usepackage{bm}

\usepackage{cite}
\usepackage[dvips]{graphicx}
\usepackage{color}

\makeatletter
\@addtoreset{equation}{section}
\makeatother

\begin{document} 

\begin{titlepage}

\baselineskip 10pt
\hrule 
\vskip 5pt
\leftline{}
\leftline{Chiba Univ. Preprint
          \hfill   \small \hbox{\bf CHIBA-EP-175}}
\leftline{\hfill   \small \hbox{April 2009}}
\vskip 5pt
\baselineskip 14pt
\hrule 
\vskip 1.0cm
\centerline{\Large\bf 
} 
\vskip 0.3cm
\centerline{\Large\bf  
Kugo--Ojima color confinement criterion
}
\vskip 0.3cm
\centerline{\Large\bf  
and Gribov-Zwanziger 
 horizon condition
}
\vskip 0.3cm

\vskip 0.5cm

\centerline{{\bf 
Kei-Ichi Kondo,$^{\dagger,{1},{2}}$  
}}  
\vskip 0.5cm
\centerline{\it
${}^{1}$Department of Physics, University of Tokyo,
Tokyo 113-0033, Japan
}
\centerline{\it
${}^{2}$Department of Physics, 
Chiba University, Chiba 263-8522, Japan
}
\vskip 1cm

\begin{abstract}
We rewrite the Zwanziger horizon condition in terms of the Kugo-Ojima parameter for color confinement.  This enables one to explain which value of the Kugo-Ojima parameter is allowed if the horizon condition is imposed. 
Although all the calculations are performed in the limit of vanishing Gribov parameter for simplicity, the obtained value is consistent with the result of numerical simulations.  Consequently, the ghost propagator behaves like free and the gluon propagator is non-vanishing at low momenta, in harmony with recent lattice results and decoupling solution of the Schwinger-Dyson equation.  The Kugo-Ojima criterion is realized only when the restriction is removed. 
\end{abstract}

Key words: color confinement, Kugo-Ojima, Gribov-Zwanziger, horizon function, 
 
\vskip 0.5cm

PACS: 12.38.Aw, 12.38.Lg 
\hrule  
\vskip 0.1cm
${}^\dagger$ 
On sabbatical leave of absence from Chiba University. 

  E-mail:  {\tt kondok@faculty.chiba-u.jp}

\par 
\par\noindent


\vskip 0.5cm

\pagenumbering{roman}
\tableofcontents




\end{titlepage}


\pagenumbering{arabic}

\baselineskip 14pt
\section{Introduction}

To solve color confinement problem is equal to answering a fundamental question of how the non-Abelian gauge theory can be quantized in a non-perturbative manner beyond the perturbation theory, as represented by the Gribov problem \cite{Gribov78}. 
This topic is currently under extensive studies due to a fact that color confinement is closely related to the asymptotic behavior of the ghost and gluon propagators in the deep infrared region, causing a challenge for numerical simulations on larger lattices.

It was shown \cite{Zwanziger89} that the partition function of the $D$-dimensional Euclidean Yang-Mills theory restricted to the first Gribov region for avoiding Gribov copies can be written in the form: 
\begin{equation}
 Z_{\gamma} := \int \mathcal{D}\mathscr{A} \delta (\partial^\mu \mathscr{A}_\mu) \det M \exp \{ -S_{YM} + \gamma \int d^D x h(x) \} 
 , 
 \label{YM1}
\end{equation}
where $S_{YM}$ is the Yang-Mills action, $M$ is the Faddeev-Popov operator $M:=-\partial_\mu D_\mu=-\partial_\mu (\partial_\mu+g \mathscr{A}_\mu \times)$ and $h(x)=h[\mathscr{A}](x)$ is the Zwanziger horizon function given by  
\begin{equation}
 h(x) 
:= - \int d^Dy gf^{ABC} \mathscr{A}_\mu^{B}(x) (M^{-1})^{CE}(x,y) gf^{AFE} \mathscr{A}_\mu^{F}(y)
 . 
\end{equation}
Here the parameter $\gamma$ called the Gribov parameter is determined by solving a gap equation, commonly called the horizon condition:
\begin{equation}
 \langle h(x) \rangle^{\gamma} = (N^2-1)D .
\end{equation}
The action corresponding to the partition function (\ref{YM1}) contains the {\it non-local} horizon term: 
\begin{equation}
 \int d^Dx h(x) 
:= - \int d^Dx \int d^Dy gf^{ABC} \mathscr{A}_\mu^{B}(x) (M^{-1})^{CE}(x,y) gf^{AFE} \mathscr{A}_\mu^{F}(y)
 . 
\end{equation}

Later, it has been shown \cite{Zwanziger92,Zwanziger93} that the non-local action (\ref{YM1}) can be put in an equivalent {\it local} form by introducing a set of complex conjugate commuting variables and anticommuting ones, which is called the Gribov-Zwanziger (GZ) action. 
We do not write it explicitly, since we do not use it in this paper. 
The GZ theory is renormalizable to all orders of perturbation theory. Hence, the restriction to the (first) Gribov region $\Omega$ makes perfect sense at the quantum level, and finite results are obtained consistent with the renormalization group. 

Note that $\Omega$ is obtained as the {\it local} minima of the gauge-fixing functional 
\begin{equation}
 F := \int d^Dx \mathscr{A}_\mu^\omega(x) \mathscr{A}_\mu^\omega(x) .
\end{equation}
However, it is known that $\Omega$ still contain Gribov copies, since there exist many local minima starting from the same $\mathscr{A}$. 
The gauge field configurations that are {\it absolute} minima of the gauge-fixing functional $F$ are known as the fundamental modular region (FMR) $\Lambda_{FMR}$. However, it is extremely difficult to deal with the Yang-Mills theory by discriminating between $\Lambda_{FMR}$ and $\Omega$.

The Kugo-Ojima color confinement criterion \cite{KO79,Kugo95,Hata82}, i.e., a sufficient condition for color confinement, is given by
\begin{equation}
 u^{AB}(0) = \delta^{AB} u(0) = - \delta^{AB} 
 , i.e., u(0) = -1 , 
\end{equation}
where the so-called Kugo-Ojima parameter $u(0)$ is defined by the infrared (IR) limit $k^2 \rightarrow 0$ of the two-point function of composite operators:
\begin{equation}
 \left( g_{\mu\nu} - \frac{k_\mu k_\nu}{k^2} \right) \delta^{AB} u(k^2) =  \int d^Dx e^{ik(x-y)} \langle 0| T[(D_\mu \mathscr{C})^{A}(x) (g\mathscr{A}_\nu \times \bar{\mathscr{C}})^{B}(y) ] |0 \rangle 
  . 
\end{equation}
However, it should be noted that the Kugo-Ojima color confinement criterion was obtained in the framework of the BRST quantization for the usual Faddeev-Popov approach, which corresponds to the $\gamma=0$ case of the above Gribov-Zwanziger formulation.
 In this paper, we consider the $D$-dimensional Yang-Mills theory in the covariant gauge with a gauge fixing parameter $\alpha$ defined by
\begin{equation}
 Z  := \int [d\mathscr{A}] [d\mathscr{B}][d\mathscr{C}][d\bar {\mathscr{C}}]    \exp \{ iS_{YM}^{tot}  \} 
 , 
 \label{YM0}
\end{equation}
where
\begin{align}
 S_{YM}^{tot}  :=& S_{YM} + S_{GF+FP} ,
 \nonumber\\
 S_{YM} :=& - \int d^Dx \frac14 \mathscr{F}_{\mu\nu}  \cdot \mathscr{F}_{\mu\nu} ,
 \nonumber\\
 S_{GF+FP} :=& \int d^Dx  \left\{ \mathscr{B} \cdot \partial_\mu \mathscr{A}_\mu +  \frac{\alpha}{2} \mathscr{B} \cdot \mathscr{B} +i \bar {\mathscr{C}} \cdot \partial_\mu D_\mu \mathscr{C} \right\} ,
  \nonumber\\
 \mathscr{F}_{\mu\nu} :=& \partial_\mu \mathscr{A}_\nu - \partial_\nu \mathscr{A}_\mu +g \mathscr{A}_\mu \times \mathscr{A}_\nu ,
  \nonumber\\
 D_\mu \mathscr{C} :=& (\partial_\mu + g \mathscr{A}_\mu \times )  \mathscr{C}
 ,
\end{align}
and the dot and the cross are defined as 
\begin{equation}
  \mathscr{A} \cdot \mathscr{B} := \mathscr{A}^A \mathscr{B}^A , \quad 
  (\mathscr{A} \times \mathscr{B})^A := f^{ABC} \mathscr{A}^B \mathscr{B}^C 
  ,
\end{equation}
using the structure constant of the gauge group $G=SU(N)$.
The Landau gauge corresponds to $\alpha=0$. 
We assume that the Euclidean result is obtained by the Wick rotation of the Minkowski one. 

The usual Faddeev-Popov approach does not take care of the Gribov copy problem. 
Therefore, if one begins to avoid the Gribov copy by restricting the space of gauge field configuration, it may happen that the Kugo-Ojima criterion $u(0)=-1$ based on the Faddeev-Popov approach does not necessarily hold.  
In fact, the direct measurements on a lattice \cite{FN07} confirm $u(0) = -0.6 \sim -0.8$.  To my knowledge, there is no theoretical explanation for this result.

In this paper we discuss a relationship between the Zwanziger horizon condition and the Kugo-Ojima color confinement criterion.  In other words, we study how the Zwanziger horizon condition imposes a constraint on the possible value of the Kugo-Ojima parameter. 
For this purpose, we give an estimation of the average of the horizon function $ \langle h(x) \rangle^{\gamma}$ in terms of the Kugo-Ojima parameter $u(0)$. 
We obtain some relations connecting three body and four body two-point functions for gluon and ghost fields, which able to give somewhat stronger results than those obtained from the Schwinger-Dyson equations for the ghost propagator and the ghost-antighost-gluon vertex function.

However, we restrict the calculation to the $\gamma=0$ limit, i.e., in the usual Faddeev-Popov approach on which the Kugo-Ojima is actually based. 
The $\gamma \ne 0$ case will be treated in a subsequent paper. 
Instead, various relations are obtained at arbitrary momentum $k$ and the deep infrared (IR) limit $k \rightarrow 0$ will be taken only in the final stage. 
Finally, we give an argument supporting that the result obtained in the limit $\gamma=0$ is not far from the correct value.

%
\section{Two-point functions of composite operators in a covariant gauge}

In this paper, we consider the two-point function for composite operators.  We use an abbreviated notation for the  Fourier transform: 
\begin{equation}
  \langle \phi_1^A \phi_2^B \rangle_{k} := 
\int d^4x e^{ik(x-y)} \langle 0| T[\phi_1^A(x) \phi_2^B(y) ] |0 \rangle 
 . 
\end{equation} 
In particular, we pay attention to the deep infrared limit $k \rightarrow 0$.
We begin with reproducing the result of Kugo \cite{Kugo95}.

First, we consider the three-body two-point function, 
$
\langle (g \mathscr{A}_\mu \times \mathscr{C})^A \bar{\mathscr{C}}^B \rangle_k
$.
Using the definition of the covariant derivative, 
\begin{equation}
 (D_\mu \mathscr{C})^A(x) := \partial_\mu \mathscr{C}^A(x) +  (g \mathscr{A}_\mu \times \mathscr{C})^A(x) 
 = \partial_\mu \mathscr{C}^A(x) +  g f^{ABC} \mathscr{A}_\mu^B(x) \mathscr{C}^C(x) 
 , 
\end{equation} 
we obtain
\begin{align}
 \langle (g \mathscr{A}_\mu \times \mathscr{C})^A \bar{\mathscr{C}}^B \rangle_k
 =& \langle (D_\mu \mathscr{C})^A \bar{\mathscr{C}}^B \rangle_k - \langle \partial_\mu \mathscr{C}^A \bar{\mathscr{C}}^B \rangle_k 
\nonumber\\
=&  i\frac{k_\mu}{k^2} \delta^{AB} + ik_\mu \langle   \mathscr{C}^A \bar{\mathscr{C}}^B \rangle_k 
\nonumber\\
=&  i k_\mu \left( \frac{1}{k^2} \delta^{AB} +   \langle   \mathscr{C}^A \bar{\mathscr{C}}^B \rangle_k \right) 
 ,
 \label{f1}
\end{align}
where we have used in the second equality the integration by parts
\begin{equation}
 \partial_\mu^x \rightarrow -ik_\mu
  ,
 \label{id0}
\end{equation} 
and a well-known relation \cite{KO79,Kugo95}
\begin{equation}
 \langle (D_\mu \mathscr{C})^A \bar{\mathscr{C}}^B \rangle_k
 = i\frac{k_\mu}{k^2} \delta^{AB} 
 .
 \label{id1}
\end{equation} 
It is very useful to introduce the one-particle irreducible (1PI) part by
\begin{equation}
 \langle (g \mathscr{A}_\mu \times \mathscr{C})^A \bar{   \mathscr{C}}^B \rangle_k^{1PI} 
:= \langle (g \mathscr{A}_\mu \times \mathscr{C})^A \bar{\mathscr{C}}^B \rangle_k/\langle   \mathscr{C}^A \bar{\mathscr{C}}^B \rangle_k 
 .
\label{def2}
\end{equation} 
Similarly, we can define another 1PI part:
\begin{align}
 \langle \mathscr{C}^A (g \mathscr{A}_\nu \times \bar{   \mathscr{C}})^B \rangle_k^{1PI} 
 =\langle   \mathscr{C}^A (g \mathscr{A}_\nu \times  \bar{\mathscr{C}})^B \rangle_k/\langle   \mathscr{C}^A \bar{\mathscr{C}}^B \rangle_k 
  .
 \label{f2b}
\end{align}
Then the above result (\ref{f1}) is also written in the form
\begin{equation}
 \langle (g \mathscr{A}_\mu \times \mathscr{C})^A \bar{   \mathscr{C}}^B \rangle_k^{1PI} 
= -i k_\mu \left( -1 + \frac{-1}{k^2} \langle   \mathscr{C}^A \bar{\mathscr{C}}^B \rangle_k^{-1}   \right) 
 .
\label{f2}
\end{equation} 
This result (\ref{f1}) or (\ref{f2}) is stronger than the result of the naive analysis of the SD equation for the ghost propagator, since only the contracted form 
$k_\mu \langle (g \mathscr{A}_\mu \times \mathscr{C})^A \bar{   \mathscr{C}}^B \rangle_k^{1PI}$
 appears in the SD equation for the ghost propagator in momentum space:   
\begin{subequations}
\begin{equation}
 \delta^{AB} = -k^2 \langle   \mathscr{C}^A \bar{\mathscr{C}}^B \rangle_k 
-i k^\mu \langle (g \mathscr{A}_\mu \times \mathscr{C})^A \bar{   \mathscr{C}}^B \rangle_k 
 ,
\label{SD1}
\end{equation} 
or
\begin{equation}
 \langle   \mathscr{C}^A \bar{\mathscr{C}}^B \rangle_k^{-1} = -k^2  \delta^{AB} 
-i k^\mu \langle (g \mathscr{A}_\mu \times \mathscr{C})^A \bar{   \mathscr{C}}^B \rangle_k^{1PI} 
 .
\label{SD2}
\end{equation}
\end{subequations}
In fact, (\ref{f1}) or (\ref{f2}) automatically satisfies this SD equation.

Next, we consider the four-body two-point function, 
$
\langle   (g \mathscr{A}_\mu \times \mathscr{C})^A (g \mathscr{A}_\nu \times  \bar{\mathscr{C}})^B \rangle_k
$.
It is also well known \cite{KO79,Kugo95}
that the following identify holds.
\begin{align}
 0 = \langle   (\partial^\mu D_\mu \mathscr{C})^A (g \mathscr{A}_\nu \times  \bar{\mathscr{C}})^B \rangle_k
 ,
 \label{f3}
\end{align}
or 
\begin{align}
 0 = \langle   (\partial^\mu \partial_\mu \mathscr{C})^A (g \mathscr{A}_\nu \times  \bar{\mathscr{C}})^B \rangle_k
 + \langle   \partial^\mu (g \mathscr{A}_\mu \times \mathscr{C})^A (g \mathscr{A}_\nu \times  \bar{\mathscr{C}})^B \rangle_k
  .
 \label{f3b}
\end{align}
Therefore, we have
\begin{align}
 ik^\mu  \langle   (g \mathscr{A}_\mu \times \mathscr{C})^A (g \mathscr{A}_\nu \times  \bar{\mathscr{C}})^B \rangle_k
 = -k^2 \langle   \mathscr{C}^A (g \mathscr{A}_\nu \times  \bar{\mathscr{C}})^B \rangle_k
  .
 \label{f4}
\end{align}

We now define the modified 1PI (m1PI) part of
$
\langle   (g \mathscr{A}_\mu \times \mathscr{C})^A (g \mathscr{A}_\nu \times  \bar{\mathscr{C}})^B \rangle_k
$ 
by
\begin{align}
& \langle   (g \mathscr{A}_\mu \times \mathscr{C})^A (g \mathscr{A}_\nu \times  \bar{\mathscr{C}})^B \rangle_k^{m1PI}
\nonumber\\
:=& 
\langle   (g \mathscr{A}_\mu \times \mathscr{C})^A (g \mathscr{A}_\nu \times  \bar{\mathscr{C}})^B \rangle_k
-  \langle (g \mathscr{A}_\mu \times \mathscr{C})^A \bar{   \mathscr{C}}^C \rangle_k^{1PI} 
 \langle \mathscr{C}^C \bar{\mathscr{C}}^D \rangle_k 
 \langle \mathscr{C}^D (g \mathscr{A}_\nu \times \bar{   \mathscr{C}})^B \rangle_k^{1PI} 
  .
\label{def3}  
\end{align}

The 1PI part should be defined from the connected part.  Hence, in the above definition, $\langle   (g \mathscr{A}_\mu \times \mathscr{C})^A (g \mathscr{A}_\nu \times  \bar{\mathscr{C}})^B \rangle_k$ must be replaced by $\langle   (g \mathscr{A}_\mu \times \mathscr{C})^A (g \mathscr{A}_\nu \times  \bar{\mathscr{C}})^B \rangle_k^{conn}
:=\langle   (g \mathscr{A}_\mu \times \mathscr{C})^A (g \mathscr{A}_\nu \times  \bar{\mathscr{C}})^B \rangle_k
- \int \frac{d^D p}{(2\pi)^D} gf^{ABC} gf^{ADE} \langle  \mathscr{A}_\mu^B   \mathscr{A}_\nu^D \rangle_{k-p}
\langle   \mathscr{C}^C   \bar{\mathscr{C}}^E \rangle_p$. 
That is to say, 
\begin{align}
\nonumber\\
&  \langle   (g \mathscr{A}_\mu \times \mathscr{C})^A (g \mathscr{A}_\nu \times  \bar{\mathscr{C}})^B \rangle_k^{conn}
\nonumber\\
:=& \langle   (g \mathscr{A}_\mu \times \mathscr{C})^A (g \mathscr{A}_\nu \times  \bar{\mathscr{C}})^B \rangle_k
- \int \frac{d^D p}{(2\pi)^D} gf^{ABC} gf^{ADE} \langle  \mathscr{A}_\mu^B   \mathscr{A}_\nu^D \rangle_{k-p}
\langle   \mathscr{C}^C   \bar{\mathscr{C}}^E \rangle_p
\nonumber\\
=&    \langle (g \mathscr{A}_\mu \times \mathscr{C})^A \bar{   \mathscr{C}}^C \rangle_k^{1PI} 
 \langle \mathscr{C}^C \bar{\mathscr{C}}^D \rangle_k 
 \langle \mathscr{C}^D (g \mathscr{A}_\nu \times \bar{   \mathscr{C}})^B \rangle_k^{1PI}
 + \langle   (g \mathscr{A}_\mu \times \mathscr{C})^A (g \mathscr{A}_\nu \times  \bar{\mathscr{C}})^B \rangle_k^{1PI}
 .
\label{def4}  
\end{align}
Therefore, 1PI and m1PI is related as 
\begin{align}
 &  \langle   (g \mathscr{A}_\mu \times \mathscr{C})^A (g \mathscr{A}_\nu \times  \bar{\mathscr{C}})^B \rangle_k^{m1PI}
\nonumber\\
=& \langle   (g \mathscr{A}_\mu \times \mathscr{C})^A (g \mathscr{A}_\nu \times  \bar{\mathscr{C}})^B \rangle_k^{1PI}
  + \int \frac{d^D p}{(2\pi)^D} gf^{ABC} gf^{ADE} \langle  \mathscr{A}_\mu^B   \mathscr{A}_\nu^D \rangle_{k-p}
\langle   \mathscr{C}^C   \bar{\mathscr{C}}^E \rangle_p
   .
\label{def5}  
\end{align}
In what follows, we use m1PI part rather than 1PI one. 

By using (\ref{f4}) and (\ref{f1}), then, it turns out that 
\begin{align}
& ik^\mu  \langle   (g \mathscr{A}_\mu \times \mathscr{C})^A (g \mathscr{A}_\nu \times  \bar{\mathscr{C}})^B \rangle_k^{m1PI}
\nonumber\\
 =& -k^2 \langle   \mathscr{C}^A (g \mathscr{A}_\nu \times  \bar{\mathscr{C}})^B \rangle_k
 -  ik^\mu \langle (g \mathscr{A}_\mu \times \mathscr{C})^A \bar{   \mathscr{C}}^D \rangle_k 
 \langle \mathscr{C}^D (g \mathscr{A}_\nu \times \bar{   \mathscr{C}})^B \rangle_k^{1PI} 
\nonumber\\
 =& -k^2 \langle   \mathscr{C}^A (g \mathscr{A}_\nu \times  \bar{\mathscr{C}})^B \rangle_k
 -    \left( - \delta^{AD} - k^2  \langle   \mathscr{C}^A \bar{\mathscr{C}}^D \rangle_k \right)
 \langle \mathscr{C}^D (g \mathscr{A}_\nu \times \bar{   \mathscr{C}})^B \rangle_k^{1PI} 
\nonumber\\
 =&
 \langle \mathscr{C}^A (g \mathscr{A}_\nu \times \bar{   \mathscr{C}})^B \rangle_k^{1PI}  
 ,
 \label{f5}
\end{align}  
where we have used the definition of 
$
\langle (g \mathscr{A}_\mu \times \mathscr{C})^A \bar{   \mathscr{C}}^D \rangle_k^{1PI}
$  
and 
$
 \langle \mathscr{C}^A (g \mathscr{A}_\nu \times \bar{   \mathscr{C}})^B \rangle_k^{1PI} 
$. 
Thus we have obtained an important relationship connecting  the three-body two-point  function and the four-body one:
\begin{align}
  \langle \mathscr{C}^A (g \mathscr{A}_\nu \times \bar{   \mathscr{C}})^B \rangle_k^{1PI}  
= ik_\mu  \langle   (g \mathscr{A}_\mu \times \mathscr{C})^A (g \mathscr{A}_\nu \times  \bar{\mathscr{C}})^B \rangle_k^{m1PI}
 .
 \label{rel-1}
\end{align} 
This relation was already pointed out to hold in Kugo \cite{Kugo95} by a diagrammatical consideration (without the explicit derivation). 
This relation should be compared with (\ref{f4}).

In order to  extract more property of the 1PI part of the four-body function, we note that it is also written as
\begin{align}
&  \langle   (g \mathscr{A}_\mu \times \mathscr{C})^A (g \mathscr{A}_\nu \times  \bar{\mathscr{C}})^B \rangle_k^{m1PI}
\nonumber\\
=&     \langle   (D_\mu \mathscr{C})^A (g \mathscr{A}_\nu \times  \bar{\mathscr{C}})^B \rangle_k 
- \langle (D_\mu \mathscr{C})^A  \bar{   \mathscr{C}}^C \rangle_k
\langle \mathscr{C}^C (g \mathscr{A}_\nu \times \bar{   \mathscr{C}})^B \rangle_k^{1PI}  
 ,
 \label{f7}
\end{align}  
since the definition (\ref{def3}) leads to 
\begin{align}
& \langle   (g \mathscr{A}_\mu \times \mathscr{C})^A (g \mathscr{A}_\nu \times  \bar{\mathscr{C}})^B \rangle_k^{m1PI}
\nonumber\\
=& 
\langle   (D_\mu \mathscr{C})^A (g \mathscr{A}_\nu \times  \bar{\mathscr{C}})^B \rangle_k
- 
\langle   (\partial_\mu \mathscr{C})^A (g \mathscr{A}_\nu \times  \bar{\mathscr{C}})^B \rangle_k
\nonumber\\
&-  \langle (g \mathscr{A}_\mu \times \mathscr{C})^A \bar{   \mathscr{C}}^D \rangle_k
 \langle \mathscr{C}^D (g \mathscr{A}_\nu \times \bar{   \mathscr{C}})^B \rangle_k^{1PI} 
\nonumber\\
=& 
\langle   (D_\mu \mathscr{C})^A (g \mathscr{A}_\nu \times  \bar{\mathscr{C}})^B \rangle_k
- 
\langle   (\partial_\mu \mathscr{C})^A (g \mathscr{A}_\nu \times  \bar{\mathscr{C}})^B \rangle_k
\nonumber\\
&-  \langle (D_\mu \mathscr{C})^A \bar{   \mathscr{C}}^C \rangle_k
 \langle \mathscr{C}^C (g \mathscr{A}_\nu \times \bar{   \mathscr{C}})^B \rangle_k^{1PI} 
 +  \langle (\partial_\mu \mathscr{C})^A \bar{   \mathscr{C}}^C \rangle_k
 \langle \mathscr{C}^C (g \mathscr{A}_\nu \times \bar{   \mathscr{C}})^B \rangle_k^{1PI}
\nonumber\\
=& 
\langle   (D_\mu \mathscr{C})^A (g \mathscr{A}_\nu \times  \bar{\mathscr{C}})^B \rangle_k
- (-i k_\mu)
\langle   \mathscr{C}^A (g \mathscr{A}_\nu \times  \bar{\mathscr{C}})^B \rangle_k
\nonumber\\
&-  \langle (D_\mu \mathscr{C})^A \bar{   \mathscr{C}}^C \rangle_k
 \langle \mathscr{C}^C (g \mathscr{A}_\nu \times \bar{   \mathscr{C}})^B \rangle_k^{1PI} 
 +  (-i k_\mu) \langle  \mathscr{C}^A \bar{   \mathscr{C}}^C \rangle_k
 \langle \mathscr{C}^C (g \mathscr{A}_\nu \times \bar{   \mathscr{C}})^B \rangle_k^{1PI}
  .
\label{def8}  
\end{align} 
By using (\ref{id1}) and (\ref{rel-1}), therefore,  (\ref{f7}) is cast into
\begin{align}
&  \langle   (D_\mu \mathscr{C})^A (g \mathscr{A}_\nu \times  \bar{\mathscr{C}})^B \rangle_k 
\nonumber\\
=& \langle   (g \mathscr{A}_\mu \times \mathscr{C})^A (g \mathscr{A}_\nu \times  \bar{\mathscr{C}})^B \rangle_k^{m1PI}  
+  
  \langle (D_\mu \mathscr{C})^A  \bar{   \mathscr{C}}^C \rangle_k
\langle \mathscr{C}^C (g \mathscr{A}_\nu \times \bar{   \mathscr{C}})^B \rangle_k^{1PI}  
\nonumber\\
=&  \delta_{\mu}{}^{\rho}  \langle   (g \mathscr{A}_\rho \times \mathscr{C})^A (g \mathscr{A}_\nu \times  \bar{\mathscr{C}})^B \rangle_k^{m1PI} 
+ i\frac{k_\mu}{k^2} ik^\rho \langle   (g \mathscr{A}_\rho \times \mathscr{C})^A (g \mathscr{A}_\nu \times  \bar{\mathscr{C}})^B \rangle_k^{m1PI} 
\nonumber\\
=& \left( \delta_{\mu}{}^{\rho} - \frac{k_\mu k^\rho}{k^2} \right) 
\langle   (g \mathscr{A}_\rho \times \mathscr{C})^A (g \mathscr{A}_\nu \times  \bar{\mathscr{C}})^B \rangle_k^{m1PI} 
  .
 \label{f9}
\end{align}  
 Thanks to (\ref{f3}),  we can introduce the function $u(k^2)$ according to Kugo \& Ojima \cite{KO79}:
\begin{equation}
  \langle  (D_\mu \mathscr{C})^{A}  (g\mathscr{A}_\nu \times \bar{\mathscr{C}})^{B}  \rangle_{k} 
:= \left( g_{\mu\nu} - \frac{k_\mu k_\nu}{k^2} \right) \delta^{AB} u(k^2)   
 , 
 \label{KOp}
\end{equation}
or  combining this with (\ref{id1}) yields
\begin{equation}
  \langle  (D_\mu \mathscr{C})^{A} (D_\nu  \bar{\mathscr{C}})^{B}  \rangle_{k} 
= \left( g_{\mu\nu} - \frac{k_\mu k_\nu}{k^2} \right) \delta^{AB} u(k^2) - \frac{k_\mu k_\nu}{k^2} \delta^{AB} 
 . 
 \label{KOp2}
\end{equation} 
If we combine (\ref{f9}) with (\ref{KOp}), therefore, 
$
\langle   (g \mathscr{A}_\mu \times \mathscr{C})^A (g \mathscr{A}_\nu \times  \bar{\mathscr{C}})^B \rangle_k^{m1PI}
$ 
must have the form:
\begin{equation}
\langle   (g \mathscr{A}_\mu \times \mathscr{C})^A (g \mathscr{A}_\nu \times  \bar{\mathscr{C}})^B \rangle_k^{m1PI}
=  [g_{\mu\nu} u(k^2) + k_\mu k_\nu v(k^2)] \delta^{AB} 
 , 
\label{rel-2}
\end{equation} 
where $v(k^2)$ is an arbitrary function and a different parameterization 
\begin{equation}
 w(k):=k^2 v(k^2) 
 , 
\end{equation}
reads 
\begin{equation}
\langle   (g \mathscr{A}_\mu \times \mathscr{C})^A (g \mathscr{A}_\nu \times  \bar{\mathscr{C}})^B \rangle_k^{m1PI}
=  \left[ g_{\mu\nu} u(k^2) + \frac{k_\mu k_\nu}{k^2} w(k^2) \right] \delta^{AB} 
 . 
\label{rel-2b}
\end{equation}

\section{Preparing the  horizon function}

The horizon function is defined only in the Landau gauge.
However, we collect the necessary information for the same type of function as the horizon function for arbitrary covariant gauge.

The vacuum expectation value (VEV) of the would-be horizon function is 
\begin{subequations}
\begin{align}
 \langle h(x) \rangle 
&= - \int d^Dy  \langle gf^{ABC} \mathscr{A}_\mu^{B}(x) (M^{-1})^{CE}(x,y) gf^{ADE} \mathscr{A}_\mu^{D}(y) \rangle 
\\
&= - \int d^Dy  \langle gf^{ABC} \mathscr{A}_\mu^{B}(x) \mathscr{C}^{C}(x)   gf^{ADE} \mathscr{A}_\mu^{D}(y) \bar{\mathscr{C}}^{E}(y) \rangle 
\\
&= - \int d^Dy  \langle (g \mathscr{A}_\mu \times \mathscr{C})^{A} (x)   (g \mathscr{A}_\mu \times \bar{\mathscr{C}})^{A}(y) \rangle 
  , 
\end{align}
\end{subequations}
where we have used the identity which holds for any functional $ f(\mathcal{A})$ of $\mathcal{A}$:
\begin{equation}
 \langle  f(\mathcal{A})  \mathscr{C}^{A}(x)  \bar{\mathscr{C}}^{B}(y)  \rangle 
=   \langle f(\mathcal{A}) (M^{-1})^{AB}(x,y) \rangle 
 . 
\end{equation} 
Therefore, the average of the VEV of the would-be horizon function reads 
\begin{align}
 V_{D}^{-1} \int d^Dx \langle h(x) \rangle 
=& V_{D}^{-1} \int d^Dx \int d^Dy  \langle (g \mathscr{A}_\mu \times \mathscr{C})^{A} (x)   (g \mathscr{A}_\mu \times \bar{\mathscr{C}})^{A}(y) \rangle 
\nonumber\\
=& \langle  (g \mathscr{A}_\mu \times \mathscr{C})^{A}     (g \mathscr{A}_\mu \times \bar{\mathscr{C}})^{A}  \rangle_{k=0} 
  , 
\end{align}
where $V_{D}$ is the volume of the $D$-dimensional spacetime. 

The use of (\ref{f9}) and (\ref{rel-1}) yields 
\begin{align}
&  \langle   (g \mathscr{A}_\mu \times \mathscr{C})^A (g \mathscr{A}_\nu \times  \bar{\mathscr{C}})^B \rangle_k
\nonumber\\
=&     \langle   (D_\mu \mathscr{C})^A (g \mathscr{A}_\nu \times  \bar{\mathscr{C}})^B \rangle_k 
- \langle   (\partial_\mu \mathscr{C})^A (g \mathscr{A}_\nu \times  \bar{\mathscr{C}})^B \rangle_k
\nonumber\\
=& 
\left( \delta_{\mu}{}^{\rho} - \frac{k_\mu k^\rho}{k^2} \right) 
\langle   (g \mathscr{A}_\rho \times \mathscr{C})^A (g \mathscr{A}_\nu \times  \bar{\mathscr{C}})^B \rangle_k^{m1PI} 
+ ik_\mu \langle \mathscr{C}^A   \bar{\mathscr{C}}^C \rangle_k
\langle   \mathscr{C}^C (g \mathscr{A}_\nu \times  \bar{\mathscr{C}})^B \rangle_k^{1PI} 
\nonumber\\
=& 
\left( \delta_{\mu}{}^{\rho} - \frac{k_\mu k^\rho}{k^2} - k_\mu  k^\rho \langle \mathscr{C}   \bar{\mathscr{C}} \rangle_k \right) 
\langle   (g \mathscr{A}_\rho \times \mathscr{C})^A (g \mathscr{A}_\nu \times  \bar{\mathscr{C}})^B \rangle_k^{m1PI} 
\\
=& 
\left( \delta_{\mu\nu} - \frac{k_\mu k_\nu}{k^2} - k_\mu  k_\nu \langle \mathscr{C}   \bar{\mathscr{C}} \rangle_k \right) u(k^2) \delta^{AB}
-k_\mu k_\nu \langle \mathscr{C}   \bar{\mathscr{C}} \rangle_k k^2  v(k^2)  \delta^{AB}
  ,
 \label{f10}
\end{align}  
where we have used (\ref{rel-2}) in the last equality and we assumed that color symmetry is not broken so that 
\begin{equation}
\langle \mathscr{C}^A   \bar{\mathscr{C}}^B \rangle_k =   \langle \mathscr{C}    \bar{\mathscr{C}}  \rangle_k \delta^{AB}
 .
\end{equation}
Performing the Lorentz and color contractions, we obtain  
\begin{align}
&  \langle   (g \mathscr{A}_\mu \times \mathscr{C})^A (g \mathscr{A}_\mu \times  \bar{\mathscr{C}})^A \rangle_k
\nonumber\\
=& (N^2-1) 
 [ \left( D-1 - k^2 \langle \mathscr{C}   \bar{\mathscr{C}} \rangle_k \right) u(k^2)  
- k^2  \langle \mathscr{C}   \bar{\mathscr{C}} \rangle_k  k^2  v(k^2)   ]
\nonumber\\
=& (N^2-1) 
 \left\{ \left( D-1 \right) u(k^2) - k^2 \langle \mathscr{C}   \bar{\mathscr{C}} \rangle_k  [u(k^2)  + k^2  v(k^2)] \right\}
 .
 \label{f11}
\end{align}  
Therefore we arrive at the average of the would-be horizon function written in terms of the Kugo-Ojima parameter and the ghost propagator at $k=0$:
\begin{align}
 \langle  h(0) \rangle
 =& V_D^{-1} \int d^Dx \langle   h(x) \rangle
\nonumber\\
=& - \lim_{k^2 \rightarrow 0} \langle   (g \mathscr{A}_\mu \times \mathscr{C})^A (g \mathscr{A}_\mu \times  \bar{\mathscr{C}})^A \rangle_{k}
\nonumber\\
=& - (N^2-1) 
  \left\{ (D-1)u(0) - \lim_{k^2 \rightarrow 0} [k^2 \langle \mathscr{C}   \bar{\mathscr{C}} \rangle_k] [u(0) +w(0)]\right\}  
 ,
 \label{f12}
\end{align} 
where we have used
$
\lim_{k^2 \rightarrow 0}[ k^2  v(k^2)] 
= \lim_{k^2 \rightarrow 0}[w(k^2)] 
= w(0)
$.

It is observed that {\bf  the dressing function $Z(k^2) :=-k^2 \langle \mathscr{C}   \bar{\mathscr{C}} \rangle_k$ of the ghost, 
\begin{equation}
\lim_{k^2 \rightarrow 0} [k^2 \langle \mathscr{C}   \bar{\mathscr{C}} \rangle_k]
=\frac{(D-1)u(0)+\langle  h(0) \rangle/(N^2-1)}{u(0)+w(0)}
 ,
\end{equation}
is finite, namely, 
$\lim_{k^2 \rightarrow 0} [k^2 \langle \mathscr{C}   \bar{\mathscr{C}} \rangle_k]^{-1} \ne 0$,
 as long as $\langle  h(0) \rangle$ is finite, provided that $u(0) + w(0) \ne 0$. 
 }
 
Note that $w(0) \ne 0$ means the existence of the massless pole in the second part of 
$
\langle   (g \mathscr{A}_\mu \times \mathscr{C})^A (g \mathscr{A}_\nu \times  \bar{\mathscr{C}})^B \rangle_k^{m1PI}
$
in (\ref{rel-2b}) .
Usually, the absence of the massless pole in 
 assumed $w(0)=0$.  However, this is not quite rigorous.  
Therefore, we leave a possibility $w(0) \ne 0$ in what follows. 
  
Up to here, all the equations hold for any covariant gauge, i.e., any value of the gauge fixing parameter $\alpha$.  
In the Landau gauge $\alpha=0$, we can study the relationship between the horizon function and the ghost propagator in more detail, as shown below. 

\section{Restricting to the Landau gauge}
 
In what follows, we impose the Landau condition
$\alpha=0$.
Then we have an additional identity:
\begin{align}
 \langle (g \mathscr{A}_\mu \times \mathscr{C})^A  \bar{   \mathscr{C}}^B \rangle_k^{1PI,\alpha=0}  
 =     \langle   (g \mathscr{A}_\mu \times \mathscr{C})^A (g \mathscr{A}_\sigma \times  \bar{\mathscr{C}})^B \rangle_k^{m1PI,\alpha=0} (-ik_\sigma)
 .
 \label{rel-3}
\end{align} 
This is because the derivative factor acting on the antighost at the ghost-gluon-antighost vertex at the right end of the diagram can be transferred to act on the external ghost, since $\partial_\mu \mathscr{A}_\mu= 0$ in the Landau gauge.

 Substituting (\ref{f1}) and (\ref{rel-2}) into (\ref{rel-3}), we obtain a relationship between the Kugo-Ojima function and the ghost propagator in the Landau gauge:
\begin{align}
 Z(k^2)^{\alpha=0} := -k^2  \langle  \mathscr{C}^A  \bar{   \mathscr{C}}^B \rangle_k^{\alpha=0}  
 =     [1+u(k^2)+k^2 v(k^2)]^{-1} \delta^{AB}
 .
 \label{rel3}
\end{align} 
In the Landau gauge, therefore, we find using (\ref{f11})
\begin{align}
&  \langle   (g \mathscr{A}_\mu \times \mathscr{C})^A (g \mathscr{A}_\mu \times  \bar{\mathscr{C}})^A \rangle_k^{\alpha=0}
\nonumber\\
=& (N^2-1) 
 \left\{ \left( D-1 \right) u(k^2) - k^2 \langle \mathscr{C}   \bar{\mathscr{C}} \rangle_k^{\alpha=0}  [u(k^2)  + k^2  v(k^2)] \right\}
\nonumber\\
=& (N^2-1) 
 \left\{ \left( D-1 \right) u(k^2) + \frac{u(k^2)  + k^2  v(k^2)}{1+u(k^2)+k^2 v(k^2)} \right\} 
 ,
 \label{f16}
\end{align} 
and the average of the horizon function reads
\begin{align}
 \langle  h(0) \rangle^{\alpha=0}
 =& V_D^{-1} \int d^Dx \langle   h(x) \rangle^{\alpha=0}
\nonumber\\
=&  - \lim_{k^2 \rightarrow 0} \langle   (g \mathscr{A}_\mu \times \mathscr{C})^A (g \mathscr{A}_\mu \times  \bar{\mathscr{C}})^A \rangle_{k}^{\alpha=0}
\nonumber\\
=& - (N^2-1) 
  \left\{ (D-1)u(0) - \lim_{k^2 \rightarrow 0} [k^2 \langle \mathscr{C}   \bar{\mathscr{C}} \rangle_k^{\alpha=0}] [u(0)+w(0)] \right\}   
\nonumber\\
=& -  (N^2-1) 
  \left\{ (D-1)u(0) + \frac{u(0)+w(0)}{1+u(0)+w(0)} \right\}   
  .
 \label{f14}
\end{align} 
The same result is obtained even if we start from the original definition (\ref{def3}) by taking into account (\ref{rel-1}) and (\ref{rel-3}):
\begin{align}
& \langle   (g \mathscr{A}_\mu \times \mathscr{C})^A (g \mathscr{A}_\nu \times  \bar{\mathscr{C}})^B \rangle_k
\nonumber\\
:=& 
\langle   (g \mathscr{A}_\mu \times \mathscr{C})^A (g \mathscr{A}_\nu \times  \bar{\mathscr{C}})^B \rangle_k^{m1PI}
+  \langle (g \mathscr{A}_\mu \times \mathscr{C})^A \bar{   \mathscr{C}}^C \rangle_k^{1PI} 
 \langle \mathscr{C}^C \bar{\mathscr{C}}^D \rangle_k 
 \langle \mathscr{C}^D (g \mathscr{A}_\nu \times \bar{   \mathscr{C}})^B \rangle_k^{1PI} 
\nonumber\\
=& 
\langle   (g \mathscr{A}_\mu \times \mathscr{C})^A (g \mathscr{A}_\nu \times  \bar{\mathscr{C}})^B \rangle_k^{m1PI}
\nonumber\\
& + k_\sigma  k_\rho  \langle   (g \mathscr{A}_\mu \times \mathscr{C})^A (g \mathscr{A}_\sigma \times  \bar{\mathscr{C}})^C \rangle_k^{m1PI}  
 \langle \mathscr{C}^C \bar{\mathscr{C}}^D \rangle_k 
  \langle   (g \mathscr{A}_\rho \times \mathscr{C})^D (g \mathscr{A}_\nu \times  \bar{\mathscr{C}})^B \rangle_k^{m1PI} 
\nonumber\\
=& 
 [g_{\mu\nu} u(k^2) + k_\mu k_\nu v(k^2)]\delta^{AB}
 + k_\mu k_\nu  
\langle \mathscr{C}^A \bar{\mathscr{C}}^B \rangle_k 
 [ u(k^2) + k^2 v(k^2)]^2 
  ,
\label{def17}  
\end{align}
where we have used  (\ref{rel-2}) in the last step.
After performing Lorentz and color contraction, indeed, we obtain the same result as (\ref{f16}):
\begin{align}
&  \langle   (g \mathscr{A}_\mu \times \mathscr{C})^A (g \mathscr{A}_\mu \times  \bar{\mathscr{C}})^A \rangle_k^{\alpha=0}
\nonumber\\
=& (N^2-1) 
 \left\{ Du(k^2)+k^2 v(k^2)+ k^2 \langle \mathscr{C}  \bar{\mathscr{C}}  \rangle_k 
[ u(k^2) + k^2 v(k^2)]^2 \right\} 
\nonumber\\
=& (N^2-1) 
 \left\{ \left( D-1 \right) u(k^2) + \frac{u(k^2)  + k^2  v(k^2)}{1+u(k^2)+k^2 v(k^2)} \right\} 
 .
 \label{f26}
\end{align} 

Note that the horizon function (\ref{f14}) defined above differs from that obtained by taking the IR limit of  (\ref{KOp2}): 
\begin{align}
  -\lim_{k^2 \rightarrow 0} \langle   (D_\mu  \mathscr{C})^A (D_\mu   \bar{\mathscr{C}})^A \rangle_{k}^{\alpha=0}
= -(N^2-1) 
  \left\{ (D-1)u(0) -1 \right\}   
 ,
 \label{f25}
\end{align} 
except for a special value of $u(0)$ (and $w(0)$).
This is because the total derivatives in the Fourier transform can not be discarded in the above calculations.   Indeed, we have taken into account all the total derivative terms. Otherwise, (\ref{f25}) agreed with (\ref{f14}).\footnote{This should be compared with the claim of Dudal et al. \cite{Dudal09}.}

In the Landau gauge, we find
\begin{align}
 \langle (g \mathscr{A}_\mu \times \mathscr{C})^A \bar{   \mathscr{C}}^B \rangle_k^{1PI} 
=& i k_\mu \left( \delta^{AB} + \frac{1}{k^2} \langle   \mathscr{C}^A \bar{\mathscr{C}}^B \rangle_k^{-1}   \right) 
\nonumber\\
=& -i \langle   (g \mathscr{A}_\mu \times \mathscr{C})^A (g \mathscr{A}_\nu \times  \bar{\mathscr{C}})^B \rangle_k^{m1PI,\alpha=0}  k_\nu 
  ,
\label{f18}
\end{align} 
which is contracted with $-ik_\mu$ to give
\begin{align}
 k^2  \delta^{AB}   +   \langle   \mathscr{C}^A \bar{\mathscr{C}}^B \rangle_k^{-1} 
=& -ik_\mu \langle (g \mathscr{A}_\mu \times \mathscr{C})^A \bar{   \mathscr{C}}^B \rangle_k^{1PI}   
\nonumber\\
=& - k_\mu  \langle   (g \mathscr{A}_\mu \times \mathscr{C})^A (g \mathscr{A}_\nu \times  \bar{\mathscr{C}})^B \rangle_k^{m1PI,\alpha=0}  k_\nu 
 .
\label{f19}
\end{align}
Then the inverse ghost propagator obeys
\begin{align}
    \langle   \mathscr{C}^A \bar{\mathscr{C}}^B \rangle_k^{-1}   
=& - k^2  \delta^{AB} - k_\mu  \langle   (g \mathscr{A}_\mu \times \mathscr{C})^A (g \mathscr{A}_\nu \times  \bar{\mathscr{C}})^B \rangle_k^{m1PI,\alpha=0}  k_\nu 
\nonumber\\
=& - k_\mu \Pi_{\mu\nu}^{AB}(k)  k_\nu 
  ,
\label{f20}
\end{align}
where we have introduced 
\begin{align}
  \Pi_{\mu\nu}^{AB}(k) :=&  g_{\mu\nu} \delta^{AB}  +\langle   (g \mathscr{A}_\mu \times \mathscr{C})^A (g \mathscr{A}_\nu \times  \bar{\mathscr{C}})^B \rangle_k^{m1PI,\alpha=0}  
\nonumber\\
=& \delta^{AB} \{ [1+u(k^2)] g_{\mu\nu} + k_\mu k_\nu v(k^2) \} 
 .
\label{f21}
\end{align}
Thus we have another relationship between the IR limit of the ghost dressing function and the Kugo-Ojima parameter (up to the value of $w(0)$):
\begin{align}
\lim_{k^2 \rightarrow 0} [-k^2 \langle \mathscr{C}^A   \bar{\mathscr{C}}^B \rangle_k]^{-1} 
=&  \lim_{k^2 \rightarrow 0}  [ k_\mu \Pi_{\mu\nu}^{AB}(k)  k_\nu/k^2 ]
\nonumber\\
=&   \delta^{AB}  [1+u(0)+w(0)] 
\nonumber\\
=&    \Pi_{\mu\mu}^{AB}(0)/D + (1-1/D) w(0)
 .
\end{align}

\section{Horizon condition and Kugo-Ojima parameter}
The horizon condition is written as
\begin{equation}
 \langle h(0)\rangle^{\gamma} = (N^2-1) D
 .
\end{equation}
The actual value $\gamma^*$ of the Gribov parameter $\gamma$ is determined by solving this gap equation in a self-consistent way.
Suppose that $\gamma$ is so small that the left-hand side can be expanded in powers of $\gamma$ around $\gamma=0$:
\begin{align}
 \langle h(0)\rangle^{\gamma} 
\equiv & \frac{1}{V_D} \frac{\partial \ln Z_\gamma}{\partial \gamma} 
\nonumber\\
=& \frac{1}{V_D} \frac{\partial \ln Z_\gamma}{\partial \gamma}\Big|_{\gamma=0} + \gamma  \frac{1}{V_D} \frac{\partial^2 \ln Z_\gamma}{\partial \gamma^2}\Big|_{\gamma=0} + O(\gamma^2) 
\nonumber\\
=& \langle h(0)\rangle_{\gamma=0} + \gamma  \int d^Dy  \langle h(0); h(y) \rangle_{\gamma=0}^{\rm connect.} + O(\gamma^2) 
  . 
\end{align}
Here the expectation value at $\gamma=0$ is to be calculated in the usual Faddeev-Popov approach. 
In fact, we have obtained $\langle h(0)\rangle_{\gamma=0}$ as a function of $u(0)$ in (\ref{f14}). 
   

\begin{figure}
\begin{center}
\includegraphics[width=3.0in]{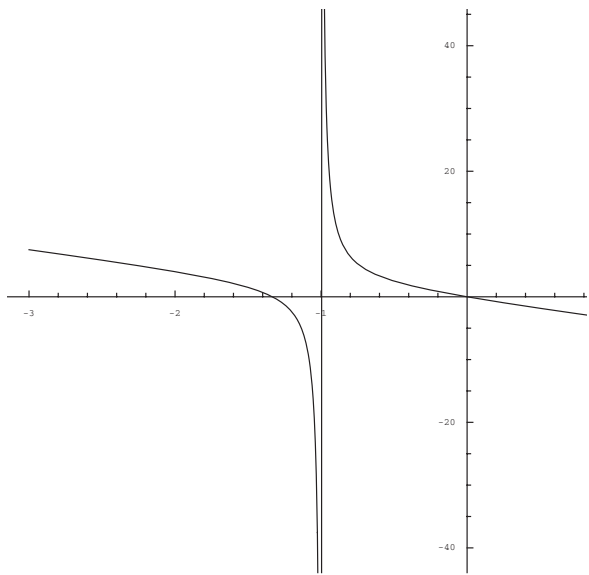}
\end{center}
\caption{
The plot of $\langle h(0)\rangle^{\gamma=0}$ versus $u(0)$ for $D=4$.
}
\label{fig:horizon}
\end{figure}

We consider how the Zwanziger horizon condition (of reducing  the integration region to the first Gribov region) restricts the possible value of the Kugo-Ojima parameter $u(0)$. 
If $\gamma^*$ is small, the horizon condition is approximately written by using $\langle  h(0) \rangle^{\alpha=0}_{\gamma=0}$: 
\begin{align}
 \langle  h(0) \rangle^{\alpha=0}_{\gamma=0}
  =& - (N^2-1) 
  \left\{ (D-1)u(0) + \frac{u(0)+w(0)}{1+u(0)+w(0)} \right\}    \cong (N^2-1)D 
  .
 \label{f22}
\end{align}

First, we consider the case of $w(0)=0$.  Then the horizon condition at $\gamma=0$ is 
\begin{align}
 \langle  h(0) \rangle^{\alpha=0}_{\gamma=0}
  =& - (N^2-1) 
  \left\{ (D-1)u(0) + \frac{u(0)}{1+u(0)} \right\}    \cong (N^2-1)D 
  .
 \label{f22a}
\end{align}
See Fig.~\ref{fig:horizon} for a plot of $\langle h(0)\rangle_{\gamma=0}$ versus $u(0)$ for $D=4$. 
Note that $u(0) \rightarrow 0$ in the vanishing coupling limit $g \rightarrow 0$. 
Therefore, the allowed region for $u(0)$ is $u(0) \in (-1,0]$ for $\langle  h(0) \rangle^{\alpha=0}_{\gamma=0} \in [0, +\infty]$ in a branch going through the origin $u(0)=0$.
The horizon condition (\ref{f22}) leads to the second degree algebraic equation for $u(0)$, 
$(D-1)u^2+2Du+D=0$, whose solutions are given by
$u=(-D \pm \sqrt{D})/(D-1)$
e.g., for $D=4$,  $3u^2+8u+4=0$ has solutions $u(0)=-2/3, -2$, and we adopt 
\begin{equation}
u(0)=- \frac23  ,
\end{equation}
because  $u(0)=-2/3$ belongs to a branch going through the origin, while this is not the case for  $u(0)=-2$.

Even for $\gamma=0$, the Zwanziger horizon condition 
$\langle  h(0) \rangle^{\alpha=0}=(N^2-1)D$
does not agree with the Kugo-Ojima color confinement  criterion $u(0)=-1$. 
In this approximation $\gamma^* \simeq 0$, the Kugo-Ojima   criterion $u(0)=-1$ is realized only when $\langle h(0)\rangle_{\gamma=0}=+\infty$, which implies that there is no restriction for the gauge field space. 

Within this approximation $\gamma^* \simeq 0$,  the horizon condition forces the ghost propagator to behave like free $1/k^2$ at $k=0$, no more singular than $1/k^2$: for $D=4$,  irrespective of the number of color $N$, 
\begin{align}
\lim_{k^2 \rightarrow 0} [-k^2 \langle \mathscr{C}^A   \bar{\mathscr{C}}^B \rangle_k]^{-1} 
=  \delta^{AB}  [1+u(0)]
=  \frac13   \delta^{AB} 
\ne 0 ,  \quad 
Z(0) = 3 
 .
\end{align}

Incidentally, the direct measurements of the Kugo-Ojima parameter $u(0)$ on a lattice have been performed by imposing the absolute Landau gauge which restricts the integration region to the FMR. The result \cite{FN07} is $u(0) \simeq -0.6 \sim -0.8$. 
Our rough estimate $u(0)=-2/3 \cong -0.67$ is very near to this result of numerical simulations.  
This suggests that $\gamma^*$ is relatively small. 
See \cite{Dudaletal08} for another support for this approximation.

Incidentally, the formal power series, 
$
  \frac{u(0)}{1+u(0)} = u(0)[1+u(0)]^{-1}
  = u(0) - u(0)^2 + u(0)^3 + \cdots 
$
yields the horizon condition 
\begin{align}
 \langle  h(0) \rangle^{\alpha=0}_{\gamma=0}
  =&  (N^2-1) 
  \left\{ - Du(0) + u(0)^2 - u(0)^3 + \cdots  \right\}    \cong (N^2-1)D 
  .
 \label{f23}
\end{align}
If we took into account only a linear term in $u(0)$ on the left-hand side, 
then the Kugo-Ojima criterion $u(0)=-1$  would be satisfied and the ghost dressing function $Z(0) =[1+u(0)]^{-1}$ would diverge. 

From the viewpoint of perturbation theory in the coupling constant $g$, the Kugo-Ojima parameter $u(0)$ begins with the order $g^2$, i.e., $u(0)=u_1 g^2 + u_2 g^4 + \cdots$. 
If we compare both sides of (\ref{f23}) order by order in the coupling constant, then the Kugo-Ojima criterion is satisfied $u(0)=-1$ in the lowest $O(g^2)$, and the ghost dressing function $Z(0) =[1+u(0)]^{-1}$ diverges, in agreement with the Gribov original result \cite{Gribov78} where $O(g^2)$ terms are taken into account.

Second, we consider the case of $w(0) \ne 0$. 
The horizon condition alone is not sufficient to determine both $u(0)$ and $w(0)$. 
Suppose the Kugo-Ojima confinement criterion is satisfied $u(0)=-1$.  Then the horizon condition at $\gamma=0$ is
\begin{align}
 \langle  h(0) \rangle^{\alpha=0}_{\gamma=0}
  =& - (N^2-1) 
  \left\{ -D+1 + \frac{-1+w(0)}{w(0)} \right\}    \cong (N^2-1)D 
  .
 \label{f22b}
\end{align}
This leads to the value of $w(0)$ irrespective of the spacetime dimension $D$ and the number of color $N$:
\begin{equation}
 w(0) = 1/2 \quad \text{for any} \quad D .
\end{equation} 
Even if $u(0)=-1$,  therefore, the ghost propagator  behaves like free $1/k^2$ at $k=0$, no more singular than $1/k^2$: 
irrespective of the spacetime dimension $D$ and the number of color $N$
\begin{align}
\lim_{k^2 \rightarrow 0} [-k^2 \langle \mathscr{C}^A   \bar{\mathscr{C}}^B \rangle_k]^{-1} 
=  \delta^{AB}   w(0)
=  \frac12   \delta^{AB} 
\ne 0, \quad 
Z(0) = 2
 .
\end{align}

Thus, the ghost propagator behaves like free at low momenta, while the gluon propagator is non-vanishing at low momenta.

\section{Conclusion and discussion}

In this paper, we have rewritten the Zwanziger horizon condition in terms of the Kugo-Ojima parameter $u(0)$ for color confinement.  Then we have  examined the region allowed for the Kugo-Ojima parameter under the horizon condition. 
Although all the calculations are performed in the limit of vanishing Gribov parameter for simplicity, the obtained value $u(0)=-2/3$ for $D=4$ is consistent with the result of numerical simulations \cite{FN07}.

Consequently, the ghost propagator behaves like free and the gluon propagator is non-vanishing at low momenta, in harmony with recent lattice results \cite{lattice,Maas09}, decoupling solution of the Schwinger-Dyson equation \cite{Boucaudetal08} and other approaches \cite{ABP08,Dudaletal08,Dudaletal05}.  

Our result suggests that the Kugo-Ojima criterion for color confinement $u(0)=-1$ will be realized only when the Zwanziger horizon condition is removed (or weakened) in a certain way.
In order to clarify this issue, the $\gamma \ne 0$ case \cite{BS09} will be treated in a subsequent paper. 

[Note added]
In preparing this paper, an interesting paper by Zwanziger was posted to the archive \cite{Zwanziger09}, in which the $\gamma \ne 0$ case was treated. 
The essential difference between the Zwanziger result for $\gamma \ne 0$ \cite{Zwanziger09} and ours for $\gamma=0$ comes from the 2nd term 
$
\langle (g \mathscr{A}_\mu \times \mathscr{C})^A \bar{   \mathscr{C}}^C \rangle_k^{1PI} 
 \langle \mathscr{C}^C \bar{\mathscr{C}}^D \rangle_k 
 \langle \mathscr{C}^D (g \mathscr{A}_\nu \times \bar{   \mathscr{C}})^B \rangle_k^{1PI} 
$
which is written as $\Delta_{sing}(k)$ in \cite{Zwanziger09}.
At least in the $\gamma=0$ case, namely, the usual FP approach, $\Delta_{sing}(k)$ does not vanish and remains non-zero even after taking the $k=0$ limit, as we have examined in the above analyses.  
In fact, if we neglected the contribution from 
\begin{equation}
 \Delta_{sing}(k) := \langle (g \mathscr{A}_\mu \times \mathscr{C})^A \bar{   \mathscr{C}}^C \rangle_k^{1PI} 
 \langle \mathscr{C}^C \bar{\mathscr{C}}^D \rangle_k 
 \langle \mathscr{C}^D (g \mathscr{A}_\nu \times \bar{   \mathscr{C}})^B \rangle_k^{1PI}
  ,
\end{equation}
 the average of the horizon function became  instead of (\ref{f22}) equal to
\begin{align}
 \langle  h(0) \rangle^{\alpha=0}
= - (N^2-1)   Du(0)
  ,
 \label{f30}
\end{align} 
and the horizon condition $\langle  h(0) \rangle^{\alpha=0}=(N^2-1)D$ would become equivalent to the Kugo-Ojima criterion $u(0)=-1$. 
There may exist some discrepancy between $\gamma=0$ and $\gamma \ne 0$, if both papers are correct.

\section*{Acknowledgments}
The author would like to thank Dr. Akihiro Shibata, Dr. Daniele Binosi and Dr. Andre Sternbeck for sending valuable comments and information on the manuscript of this paper. 
He is grateful to High Energy Physics Theory Group and Theoretical Hadron Physics Group in the University of Tokyo, especially, Prof. Tetsuo Hatsuda for kind hospitality extended to him on sabbatical leave.
This work is financially supported by Grant-in-Aid for Scientific Research (C) 21540256 from Japan Society for the Promotion of Science
(JSPS).


\end{document}